\documentclass[aps,prb,floatfix,twocolumn,showpacs]{revtex4}
\usepackage{amsmath,amssymb,rotating,multirow}
\usepackage{graphicx}

\begin{document}

\hsize\textwidth\columnwidth\hsize\csname@twocolumnfalse\endcsname

\title{Orbital and spin relaxation in single and coupled quantum dots}

\author{Peter Stano and Jaroslav Fabian}
\affiliation{Institute for Theoretical 
Physics, University of Regensburg, 93040 Regensburg, Germany}

\vskip1.5truecm
\begin{abstract}
Phonon-induced orbital and spin relaxation rates of single electron states in lateral single and double quantum dots are obtained numerically for realistic materials parameters. The rates are calculated as a function of magnetic field and interdot coupling, at various field and quantum dot orientations. It is found that orbital relaxation is due to deformation potential phonons at low magnetic fields, while piezoelectric phonons dominate the relaxation at high fields. Spin relaxation, which is dominated by piezoelectric phonons, in single quantum dots is highly anisotropic due to the interplay of the Bychkov-Rashba and Dresselhaus spin-orbit couplings. Orbital relaxation in double dots varies strongly with the interdot coupling due to the cyclotron effects on the tunneling energy. Spin relaxation in double dots has an additional anisotropy due to anisotropic spin hot spots which otherwise cause giant enhancement of the rate at useful magnetic fields and interdot couplings. Conditions for the absence of the spin hot spots in in-plane magnetic fields (easy passages) and perpendicular magnetic fields (weak passages) are formulated analytically for different growth directions of the underlying heterostructure. It is shown that easy passages disappear (spin hot spots reappear)  if the double dot system loses symmetry by an xy-like perturbation. 
\end{abstract}
\pacs{72.25.Rb, 73.21.La, 71.70.Ej, 03.67.Lx}
\maketitle

\section{Introduction}

The spin degree of freedom in solid state systems has a much longer memory
than orbital degrees. This fact is exploited in spin electronics \cite{zutic2004:RMP}
as well as in spin quantum computing, most notably in quantum dots \cite{loss1998:PRA}
in which electron spin provides a qubit for controlled single and two-qubit
operations \cite{loss1998:PRA, fabian2005:PRB}. The performance 
of the spin qubits is ultimately limited by spin relaxation and decoherence. At present we 
believe that general principles and mechanisms of spin relaxation and decoherence  
in quantum dots are known, while it remains to develop particular models,
understand realizations of the mechanisms as well as to perform realistic
calculations, in special cases of interest.

There appear two principal mechanisms of spin relaxation in quantum dots.
At low magnetic fields (say, tens of gauss), the relaxation proceeds via
hyperfine coupling of the electron spin with the lattice or impurity 
nuclei.\cite{erlingsson2001:PRB,erlingsson2002:PRB, coish2005:PRB} On the other
hand, at higher fields (Teslas) the relaxation is due to phonon-induced
spin-flip transitions 
\cite{khaetskii2000:PRB,khaetskii2001:PRB,falko2005:PRL,aleiner2001:PRL,%
alcalde2004:PE, destefani2004:PRB, stavrou2005:PRB,fedichkin2003:PRA, san-jose2006:CM, %
florescu2006:PRB, wang2006:CM, cheng2004:PRB, romano2005:CM}.
These are allowed due to the presence of spin-orbit coupling.
Variants of the phonon-induced spin relaxation has been proposed,
such as ripple coupling,\cite{woods2002:PRB,alcade2005:MJ} important
in very small quantum dots (10 nm), or fluctuations in spin-orbit 
parameters, important when underlying heterostructure inhomogeneities
\cite{Sherman2005:PRB} are present. A possible direct spin-phonon 
coupling\cite{khaetskii2000:PRB,khaetskii2001:PRB,calero2005:PRL}, due 
to spin-orbit modulated electron-phonon coupling, have been 
found of less importance. Phonons can also change the spin precession
rate and cause spin decoherence.\cite{semenov2003:PRL} As was 
shown in Ref.  \onlinecite{golovach2004:PRL} phonon-induced spin relaxation and
decoherence proceed on similar time scales. Another source of the 
decoherence is the fluctuation of the gate potential.\cite{hu2006:PRL} 
The experimental results 
on spin relaxation in single \cite{hanson2003:PRL,hanson2005:PRL,elzerman2004:N,fujisawa2002:N,
fujisawa2002:PHB,kroutvar2004:N} and double dots \cite{johnson2005:N,%
sasaki2005:PRL}, as well as on orbital relaxation \cite{hayashi2003:PRL,petta2004:PRL},
support the above theoretical picture.

Here we present a systematic and comprehensive investigation of phonon-induced
orbital and spin relaxation in lateral single and double quantum dots, defined
in a GaAs heterostructure. We consider the most relevant electron-phonon
couplings---the deformation potential and piezoelectric acoustic phonons,
and realistic spin-orbit couplings---the Bychkov-Rashba and the linear and
cubic Dresselhaus ones. We calculate the relaxation rates in the presence
of in-plane and perpendicular magnetic fields. Our results are numerically 
exact within the Fermi's Golden rule approximating the transition rate.
We have already reported on new anisotropy effects of spin relaxation in 
double dots, in Ref. \onlinecite{stano2005:CM}. The anisotropy arises due
to anisotropic spin hot spots, the parameter (magnetic field and
interdot coupling) regions in which a spectral crossing between
a spin up and a spin down state is lifted (producing an anti-crossing) by 
spin-orbit coupling \cite{fabian1998:PRL, fabian1999:PRL}. At these points
the spin and orbital relaxation rates are equal. In single quantum dots spin hot
spots were found in Ref. \onlinecite{bulaev2005:PRB}, while in vertical
few-electron quantum dots in Ref. \onlinecite{bagga2006:CM}. 
In lateral double dots spin hot 
spots appear at useful magnetic fields (1 T) and interdot couplings (0.1 meV),
due to the crossing of the lowest orbital antisymmetric (with respect to
the quantum dot axis) state and the Zeeman split symmetric state of the 
opposite spin. \cite{stano2005:PRB} This occurs when the tunneling energy equals the Zeeman
energy. Manipulation of interdot coupling in the presence of a magnetic
field thus in general results in a short spin lifetime. Fortunately, we
have found \cite{stano2005:CM} that the spin hot spots are absent for
certain arrangements of the double dots' axis and the orientation of
the in-plane magnetic field. In particular, if the dots are oriented along
a diagonal [on a (001) heterostructure plane], and the magnetic field
is perpendicular, the spin hot spots are absent (due to symmetry reasons) 
for any values
of spin-orbit parameters. We have proposed such a geometry, which 
corresponds to what we call ``easy passage'' \cite{stano2005:CM}, for quantum
information processing. 

The particular results of our Letter, Ref. \onlinecite{stano2005:CM},
are not repeated here. Instead we
focus on providing a unified description, both analytical and
numerical, of orbital and spin relaxation rates. We give analytical formulas
describing the trends, with respect to magnetic fields and confinement energies
of the dots, of the rates. We present the numerically calculated
orbital relaxation rates and demonstrate that they are due to the deformation
potential phonons at low magnetic fields and due to piezoelectric phonons at
high fields (at zero magnetic field orbital relaxation in a biased double
dot was studied in Ref. \onlinecite{stavrou2005:PRB}, using a two-level
model). 

As for spin relaxation, we demonstrate here the different origin of 
spin hot spots in single and double quantum dots. While in single
dots spin hot spots appear due to the Bychkov-Rashba coupling \cite{bulaev2005:PRB},
in double dots both the Bychkov-Rashba and Dresselhaus couplings 
contribute. The reasons is the different symmetry of the underlying
states in single and double dots \cite{stano2005:PRB}. Furthermore, we
classify here the conditions for the absence (or narrowing) of spin hot
spots in double dots defined in quantum wells grown in different crystallographic
directions, in which the Dresselhaus spin-orbit coupling takes on different
functional forms. We also explore the orbital effects of a perpendicular
magnetic field component---the main effect is the absence of easy passages; 
only a narrow ``weak passages'' appear instead with inhibited but finite
spin hot spots. Finally, we show that easy passages are also absent
in general asymmetric double dots, implying stringent symmetry requirements
on coupled dots for spin information processing.

The paper is organized as follows. In Sec. II we describe our model of
single and double quantum dots, derive relevant perturbations 
responsible for the spin relaxation, and
write useful expressions for orbital and spin  relaxation rates. 
In Sec. III we describe the orbital and spin relaxation relaxation in single dots
for the case of in-plane and perpendicular magnetic fields. Section IV
gives a similar description for double dots. Finally, in Sec. V we give conclusions.

\section{Model}

\subsection{Hamiltonian}

We study a two-dimensional electron gas confined in a (001) plane, spanned
along x and y directions, of a
zinc-blende
semiconductor heterostructure. An additional confinement into lateral quantum
dots is given by top gates. The single electron Hamiltonian,
in the presence of magnetic field and spin-orbit coupling, is
\begin{equation} \label{eq:hamiltonian}
H=T+V_{}+H_Z+H_{BR}+H_{D}+H_{D3}. 
\end{equation}
Here $T=\mathbf{P}^2/2m$ is the kinetic energy with the effective
electron mass $m$ and kinematic momentum $\mathbf{P} = \mathbf{p} + e\mathbf {A}
= 
-i\hbar{\nabla}+e\mathbf{A}$;  $e$ is the proton
charge and  $\mathbf{A}$ is the vector potential of the magnetic field
$\mathbf{B}$. If the magnetic field is perpendicular to the plane,
$\mathbf{B}_\perp=(0,0,B_\perp)$, we choose the vector potential as
$\mathbf{A}_\perp=(B_\perp/2)(-y,x,0)$. 
 If the field is in the plane, lying under the angle $\gamma$ relative to
$\hat{x}$,
$\mathbf{B}_\parallel=B_\parallel(\cos\gamma,\sin\gamma,0)$. 
The orbital effects of the in-plane field are inhibited---only the Zeeman interaction is
taken into account in this case. 
The position vector is denoted as $\mathbf{R}=(x,y,z)=(\mathbf{r},z)$. 
We will also find useful to introduce the 
kinematic angular momentum $\mathbf{L}=\mathbf{R}\times\mathbf{P}$.

The quantum dot geometry is defined by the confining potential
\begin{equation} \label{eq:doubledot} 
V_{}(\mathbf{r}) = \frac{1}{2} m \omega_0^2
\mathrm{min}\{(\mathbf{r}-\mathbf{d})^2,(\mathbf{r}+\mathbf{d})^2\}.
\end{equation}
The distance between the 
minima of the potential is $2d$, which is further called the interdot distance. 
The angle between $\mathbf{d}$ and $\hat{x}$ is denoted as $\delta$. 
If d=0, the potential is parabolic, $V_{} = (1/2) m \omega_0^2 r^2$, representing the single dot case with the confinement energy 
$E_0 = \hbar\omega_0$ and the confinement length $l_{0} = (\hbar/m\omega_0)^{1/2}$,
setting the energy and length scale, respectively. Both the confining potential
and the vector potential of the perpendicular magnetic field define the
effective length
$l_B=l_{0}(1+B_\perp^2e^2 l_{0}^4/4\hbar^2)^{-1/4}$.

The Zeeman energy is given by
$H_Z=(g /2)\mu_B\boldsymbol{\sigma}.\mathbf{B}$, where
$g $ is the conduction band $g$ factor, $\mu_B$ is the Bohr magneton, 
and $\boldsymbol{\sigma}$ are the Pauli matrices. To shorten the
notation, we will use a renormalized magneton $\mu = (g /2)\mu_B$. 

Spin-orbit coupling gives three important terms in confined
systems.\cite{zutic2004:RMP}
The Bychkov-Rashba Hamiltonian,\cite{rashba1960:FTT, bychkov1984:JPC}
\begin{equation}
H_{BR}=\tilde{\alpha}_{BR}\left(\sigma_x P_y-\sigma_y P_x \right)/\hbar,
\end{equation}
appears if the confinement is not symmetric in the growth direction (here $\hat{z}$).
The strength $\tilde{\alpha}_{BR}$ of the interaction can be tuned by modulating
the asymmetry by top gates. Due to the lack of spatial inversion symmetry in 
zinc-blende semiconductors, the spin-orbit interaction of conduction electrons 
takes the form of the Dresselhaus Hamiltonian,\cite{dresselhaus1955:PR} which
gives 
two terms, one linear and one cubic in momentum:\cite{dyakonov1986:FTP}
\begin{eqnarray}
H_{D}&=&\gamma_c \langle P_z^2/\hbar^2 \rangle\left(-\sigma_x P_x+\sigma_y
P_y\right)/\hbar,\\
H_{D3}&=&(\gamma_c/2)\left(\sigma_x P_x P_y^2-\sigma_y P_y
P_x^2\right)/\hbar^3+H.c.,
\end{eqnarray}
where $\gamma_c$ is a material parameter. The angular brackets in $H_{D}$ denote
quantum 
averaging in the z direction--the magnitude of $H_{D}$ depends on the
z confinement strength. We express the strengths of the linear spin-orbit
interactions in length units by $l_{BR}=\hbar^2/2 m
\tilde{\alpha}_{BR}$ and 
$l_{D}=\hbar^4/2 m \gamma_c\langle P_z^2\rangle$.

In our numerical computations we use bulk GaAs materials parameters: 
$m=0.067\,m_e$, $g =-0.44$, and $\gamma_c=27.5$ eV$\mathrm{\AA^3}$. For
the coupling of the linear Dresselhaus terms we choose $\gamma_c\langle
P_z^2\rangle/\hbar^2=4.5$ meV\AA, corresponding to 
the 11 nm thick ground state of the triangular confining
potential.\cite{sousa2003:PRB}
To agree with experimental data (see Ref. \onlinecite{stano2005:CM}) we choose for $\tilde{\alpha}_{BR}$ a value of 3.3 meV\AA, which is in
line of experimental observations\cite{miller2003:PRL, knap1996:PRB} and 
corresponds to the carrier density of 5$\times10^{11}$ cm$^{-2}$ in Ref. 
\onlinecite{rossler2003:PRB}. These values correspond to length units of $l_{BR}=1.8$ $\mu$m, and $l_D=1.3$ $\mu$m.

For a confinement length of 32 nm (used in a recent experiment \cite{elzerman2004:N})
and a perpendicular magnetic field of 1 T, one gets the following typical
magnitudes for the strengths of the contributions to the
Hamiltonian given by \eqref{eq:hamiltonian}: 1.1  meV for the confinement energy $E_0$, 
13 $\mu$eV for the Zeeman splitting, and 14, 10, and 0.8 $\mu$eV
for the linear Dresselhaus, Bychkov-Rashba, and the cubic Dresselhaus terms,
respectively.
The spin-orbit interactions are small perturbations, with strengths comparable
to the Zeeman splitting. This leads to the many orders of magnitude difference
between the orbital and spin relaxation rates.

We numerically diagonalize the full Hamiltonian \eqref{eq:hamiltonian} 
(see Ref. \onlinecite{stano2005:PRB} for further details) 
and compute the orbital and spin relaxation rates using Fermi's Golden Rule. 
We also present analytical calculations for various limiting cases.

\subsection{Perturbative eigenfunctions}

Our numerical results can be qualitatively understood from considering the
lowest order of the perturbative solution of the Hamiltonian
\eqref{eq:hamiltonian}. We assume that
spin-orbit couplings are small perturbations and
that one can solve the Schr\"odinger equation for Hamiltonian $H_0=T+V_{}+H_Z$.
First, we transform\cite{aleiner2001:PRL} the Hamiltonian to remove the linear spin-orbit
terms
\begin{equation}
H\to e^{H^{op}}H e^{-H^{op}}=H_0+H_1,
\label{eq:transformation}
\end{equation}
where
\begin{equation}
H_1=H_{D3}+H_{D}^{(2)}+H_{BR}^{(2)}+H^Z_{D}+H^Z_{BR}.
\end{equation}
The
transformation is defined by operator $H^{op}=H^{op}_{BR}+H^{op}_{D}$,
\begin{equation} 
H^{op} =(i/2 l_{BR}) (y\sigma_x-x\sigma_y)-(i/2 l_{D})(x\sigma_x-y\sigma_y), 
\label{eq:operators}
\end{equation}
Keeping only terms up to the second order in the linear spin-orbit
and Zeeman couplings and the lowest order term in the cubic Dresselhaus
coupling, the new terms of the transformed Hamiltonian are
\begin{eqnarray} 
H_{D}^{(2)}&=&[H_{D},H_{D}^{op}]/2=-(\hbar^2/4 m l_{D}^2)(1-L_z \sigma_z),
\label{eq:hamiltonian2a}\\
H_{BR}^{(2)}&=&[H_{BR},H_{BR}^{op}]/2=-(\hbar^2 / 4 m l_{BR}^2)(1+L_z \sigma_z),
\label{eq:hamiltonian2b}\\
H^Z_{D}&=&[H_Z,H^{op}_{D}]=-(\mu B_\perp /l_{D}) 
(x\sigma_y+y\sigma_x)+\nonumber\\
&&+(\mu
B_\parallel /l_{D}) \sigma_z (x\sin\gamma+y\cos\gamma),
\label{eq:hamiltonian2c}\\
H^Z_{BR}&=&[H_Z,H^{op}_{BR}]=(\mu B_\perp/l_{BR})
(y\sigma_y+x\sigma_x)-\nonumber\\
&&-(\mu
B_\parallel/l_{BR})\sigma_z(x\cos\gamma+y\sin\gamma).
\label{eq:hamiltonian2d}
\end{eqnarray}
While $H_D^{(2)}$ and $H_{BR}^{(2)}$ are transformed linear spin-orbit
terms in the absence of magnetic field, the terms $H_{D}^Z$ and 
$H_{BR}^Z$ describe the mixing of the spin-orbit and Zeeman 
interaction in the unitary transformation given by $H^{op}$; these
terms are essential for understanding spin relaxation anisotropy. 

We denote the eigenfunctions and eigenenergies of $H_0$
as $\Psi$ and $\epsilon$.
We use the standard perturbation theory for non-degenerate states
and then remove the unitary
transformation to get the approximate eigenfunctions, 
$\overline{\Psi}$, of the original Hamiltonian \eqref{eq:hamiltonian}:
\begin{equation}
\overline{\Psi}_i=e^{-H^{op}}\left(\Psi_i+\sum_{j\neq i}\frac{\langle \Psi_j
|H_1 |\Psi_i\rangle}{\epsilon_i-\epsilon_j}\Psi_j\right).\\
\label{eq:eigenfunction}
\end{equation}
For degenerate states, which normally lead to spin hot spots with 
strongly enhanced spin relaxation,
we use L\"owdin's
perturbation theory. If two eigenstates, $\Psi_i$ and $\Psi_j$, of $H_0$ are degenerate,
the corresponding perturbed states are 
\begin{equation}
\overline{\Psi}_i=e^{-H^{op}}\left(\beta_{ii}\Psi_i+\beta_{ij}\Psi_j +\sum_{k\neq
i,j}\frac{\langle \Psi_k |H_1
|\Psi_i\rangle}{\epsilon_i-\epsilon_k}\Psi_k\right).
\label{eq:eigenfunction2}
\end{equation}
The coefficients $\beta$ are the solutions of the appropriate secular equation. 
If $\epsilon_i-\epsilon_j\gg (H_1)_{ij}$, then
$\beta_{ii}\sim 1$ and $\beta_{ij}\ll 1$ -- Eq.
\eqref{eq:eigenfunction} is recovered. 
The other case, when $\epsilon_i-\epsilon_j
\lesssim (H_1)_{ij}$, describes anti-crossing---spin hot spots.
\cite{fabian1998:PRL,fabian1999:PRL} In the limiting case,
when $\epsilon_i=\epsilon_j$, we get $\beta_{ii}= 1/\surd 2= \beta_{ij} $.

The analytical solution of $H_0$ for the single dot case is
known. The eigenstates are the Fock-Darwin\cite{fock1928:ZP, darwin1931:pcps} states,
$\Psi_{n,l,\sigma}$, where $n$ is the principal quantum number, $l$ is the
orbital quantum number, and $\sigma$ is the spin. 
For the double dot case the analytical solution is not known, but for our double
dot potential the eigenfunctions can be approximated by a properly symmetrized
linear combination of Fock-Darwin functions centered at the potential
minima.\cite{stano2005:PRB} 

\subsection{Phonon-induced orbital and spin relaxation rates}

By orbital relaxation we mean the transition from the
first excited orbital state to all lower lying states.
By spin relaxation we mean the transition from the upper Zeeman split orbital 
ground state to all lower lying states (except at high magnetic fields, 
there is only one lower Zeeman split orbital ground state). 
The spin of a state $\Psi$ is
quantized in the direction of the magnetic field. However, due to the spin orbit
interactions, the perturbed states $\overline{\Psi}$ have no common spin quantization axes.
We call a state to be spin up (down) if the mean value of the spin in the
direction of the magnetic field is positive (negative). Since the spin-orbit
interactions are a small perturbation, these mean values are close to $\pm 1/2$,
except at anti-crossings.

Given the initial and final states for the transition we compute the rates by
Fermi's Golden rule, where the perturbation is the electron--phonon interaction.
The relevant terms for our GaAs system comprise deformation (${\rm df}$) and piezoelectric 
acoustic (${\rm pz}$) phonons, described by Hamiltonian terms \cite{mahan}
\begin{eqnarray} 
\label{eq:Hdf}
H^\mathrm{df}&=&\sigma_e \sum_{\mathbf{K}}
\sqrt{\frac{\hbar K}{2 \rho V c_l}}
(b_{\mathbf{K},l}+b_{-\mathbf{K},l}^\dagger)e^{i\mathbf{K}.\mathbf{R}},\\
H^\mathrm{pz}&=&-i e h_{14} \sum_{\mathbf{K},\lambda}
\sqrt{\frac{\hbar}{2 \rho V c_\lambda}}
M_\lambda
(b_{\mathbf{K},\lambda}+b_{-\mathbf{K},\lambda}^\dagger)
e^{i\mathbf{K}.\mathbf{R}}.
\end{eqnarray} 

Here the three--dimensional phonon wave vector is denoted by 
$\mathbf{K}=(k_x,k_y,k_z)=(\mathbf{k},k_z)$, and $\lambda=l,t1$, or $t2$ is
the phonon's polarization (one longitudinal and two transversal);
$\rho$ is the material density ($5.3\times
10^3$kg m$^{-3}$, for GaAs), $V$ is the volume of the crystal, $c_\lambda$ is the 
sound velocity,
($c_l=5.3\times 10^3$ m/s, $c_t=2.5\times 10^3$ m/s),
 $b^\dagger_{\mathbf{K},\lambda},b_{\mathbf{K},\lambda}$ are the
creation and annihilation phonon operators, $\sigma_e$ is the deformation
potential (7.0 eV), and $eh_{14}$ is the piezoelectric constant ($1.4\times
10^9$ eV/m). Finally, the geometrical factors $M_\lambda$ are equal to $2( k_x k_y e^\lambda_z+k_z k_x e^\lambda_y+k_y k_z e^\lambda_x)/K^2$, where
$\mathbf{e}^\lambda$ are unit polarization vectors.

Consider first the deformation potential, Eq. \ref{eq:Hdf}, in 
which only longitudinal ($\lambda=l$) phonons take part.
Using Fermi's Golden rule, a relaxation (orbital or spin) rate can be
written as
\begin{eqnarray}
\label{eq:full rate def}
\Gamma^\mathrm{df}&=&[n(E_{})+1]\gamma_\mathrm{df} E_{}^2 \int {\rm d}^2
\mathbf{k}|F(\mathbf{k})|^2 |f(k_z^l)|^2/k_z^l\\
&=&[n(E_{})+1]\gamma_\mathrm{df} (E_{}^2/l_B)  \chi_{\rm df}(\mathcal{E}_l).
\label{eq:rate def}
\end{eqnarray}
Here $E_{}$ is the energy difference between the initial and final states,
$n(E)=[\exp(E/k_BT)-1]^{-1}$ is the occupation number
of the phonon state with energy $E$ at temperature $T$ (further we use zero
temperature), $\gamma_\mathrm{df}=\sigma_e^2/8\pi^2\rho c_l^4\hbar^3$ is the
strength of the deformation electron--phonon interaction [$8.3\times10^{10}$ s$^{-1}$ nm /(meV)$^2$], 
$F(\mathbf{k})=\int \mathrm{d}^2\mathbf{r}\, 
\overline{\Psi}^\dagger_i e^{i \mathbf{k}.\mathbf{r}} 
\overline{\Psi}_f$ is the xy--overlap, and $f(k_z)=\int
\mathrm{d}z\, \psi_0^\dagger e^{i k_z z} \psi_0$ is the z--overlap, contribution 
of which can be neglected, $f(k_z)\approx 1$, if 
the energy difference $E_{}$ is much smaller than the excitation energy
in the z confinement potential.
The z component of the wave vector is given by 
$k_z^\lambda=\sqrt{\mathcal{E}_\lambda^2/l_B^2-k^2}$, where
the dimensionless parameter $\mathcal{E}_\lambda=E_{} l_B/\hbar c_\lambda$ is the ratio of
the effective length, $l_B$, and the the wavelength of the emitted phonon.
Finally, $\chi_{\rm df}(\mathcal{E}_l)$ is
an integral of the xy--overlap $F(\mathbf{k})$. Since the typical linear dimension
of a wave function is the effective length $l_B$, we express it as
\begin{equation}
\chi_{\rm df}(\mathcal{E}_l)=\int_{k_z \geq 0} {\rm d}^2 (\mathbf{k} l_B)
\frac{|F(\mathbf{k})|^2}{\sqrt{\mathcal{E}_l^2 - k^2 l_B^2}}.
\label{eq:chi} 
\end{equation}
We compute the relaxation rate $\Gamma^{\rm df}$ numerically using formula \eqref{eq:full rate def}.
However, we can gain physical insight in two important limits.
First, if the wavelength of relevant phonons is smaller than the size
of the dots, $\mathcal{E}_l \gg 1$, the
square root can be taken out from the integral and $\chi_{df}\sim
1/\mathcal{E}_l$. Physically, this means that the energy to be absorbed by the phonon
is
large and almost whole is in the z component of the phonon wave vector (phonon
is emitted almost perpendicularly to the xy--plane). Second, in the
opposite limit of $\mathcal{E}_l
\ll 1$, the integration is only in the vicinity of point $\mathbf{k}=0$. Because
of the orthogonality of the eigenfunctions the overlap
integral vanishes, $F(\mathbf{k}\to
\mathbf{0})\to 0$, and the lowest order gives $|F|^2\sim (k l_B)^2$.
This leads to the dependence of
$\chi_{\rm df}(\mathcal{E}_l)\sim \mathcal{E}_l^3$. 

Analogous expression holds for the piezoelectric interaction which 
contains contributions from longitudinal and transverse phonons. 
The relaxation rate can be written as 
\begin{equation}
\Gamma^\mathrm{pz}=[n(E_{})+1]\gamma_\mathrm{pz} (1/l_B) \chi_{pz}(\mathcal{E}),
\label{eq:rate pz}
\end{equation}
with $\gamma_\mathrm{pz}=(eh_{14})^2/8\pi^2\rho c_l^2\hbar=4\times 10^{10}$
s$^{-1}$ (note the different unit from $\gamma_{\rm df}$)
nm and $\chi_{\rm pz}(\mathcal{E})=\sum_{\lambda}(c_l^2/c_\lambda^2) \int {\rm d}^2
(\mathbf{k}
l_B)  |M_\lambda|^2 |F(\mathbf{k})|^2/k_z^\lambda l_B$. The geometrical factors,
$M_\lambda$, have no influence on the limiting expressions for $\chi_{\rm pz}$ in the 
limit $\mathcal{E} \ll 1$, where $\chi_{\rm pz}(\mathcal{E})\sim \mathcal{E}^3$. If
$\mathcal{E}\gg1$, the fact that  
$M_\lambda$ contains factors  $(k_x/K)^2$ and $k_x/K$ leads to limits $\mathcal{E}^{-5}$ and $\mathcal{E}^{-3}$ for the
longitudinal and transverse phonons, respectively. Table \ref{tab:one} summarizes 
the limiting expressions.

\begin{table}
\begin{tabular}{|c|c|c|}
\hline
&$\mathcal{E}_\lambda\gg 1$&$\mathcal{E}_\lambda\ll 1$\\
\hline
$\Gamma^\mathrm{df}$&$A^2 \gamma_\mathrm{df}E_{}^2/{l_B
\mathcal{E}_l}$&$A^2 \gamma_\mathrm{df} E_{}^2\mathcal{E}^3_l/l_B$\\
$\Gamma^\mathrm{pz}_l$&$A^2 \gamma_\mathrm{pz}/{l_B
\mathcal{E}_l^5}$&$A^2 \gamma_\mathrm{pz}\mathcal{E}^3_l/l_B$\\
$\Gamma^\mathrm{pz}_t$&$A^2 \gamma_\mathrm{pz}c_l^2/l_B
\mathcal{E}_t^3 c_t^2$&$A^2
\gamma_\mathrm{pz}\mathcal{E}^3_tc_l^2/l_B c_t^2$\\
\hline
relative&$\Gamma^\mathrm{df}\gg \Gamma^\mathrm{pz}_t\gg
\Gamma^\mathrm{pz}_l$&$\Gamma^\mathrm{pz}_t\approx (c_l/c_t)^5 
\Gamma^\mathrm{pz}_l\gg \Gamma^\mathrm{df}$\\
\hline
\end{tabular}
\caption{The relaxation rates and the relative strength of the 
contributions due to deformation ($\lambda=l$), piezoelectric 
longitudinal ($\lambda=l$), and
piezoelectric transversal ($\lambda=t$) phonons. The two limiting cases are defined by the 
ratio, $\mathcal{E}_\lambda$, of the wavelength of the emitted phonon of polarization
$\lambda$, and the effective length $l_B$.
The initial and final states are encoded into the coefficient $A$, which needs to be evaluated for specific cases.}
\label{tab:one}
\end{table}

In addition to the deformation and piezoelectric phonons, 
there are additional electron-phonon spin
dependent interactions which can lead to spin relaxation. However, 
a direct spin-phonon coupling (spin-orbit modulated electron-phonon
interaction\cite{khaetskii2000:PRB,khaetskii2001:PRB}) is 
believed to give a negligible contribution.
In very small (say, 10-20 nm, which is not our case) quantum dots spin relaxation due 
to the so called ripple mechanism\cite{woods2002:PRB} can be 
as important as the spin-orbit mechanism and should be considered.
Finally, at low magnetic fields the relaxation is believed to be dominated by the 
hyperfine interaction between the electron and nuclei of the host
material.\cite{erlingsson2001:PRB,erlingsson2002:PRB}

\section{Single dots}

In the single dot case we identify the unperturbed lower and upper Zeeman split 
orbital ground, and excited orbital
states as $\Psi_{0,0,\uparrow}$, $\Psi_{0,0,\downarrow}$, and 
$\Psi_{0,-1,\uparrow}$, respectively.\cite{stano2005:PRB}
The negative value of the g factor 
energetically favors spin up rather that spin down states.
Having opposite spin, the
perturbed ground and spin states will have a nonzero overlap due to
those perturbations in the transformed Hamiltonian \eqref{eq:transformation} 
which do not commute with the
Zeeman term. 
Therefore the xy--overlap $F(\mathbf{k})$ will be proportional to the strengths 
of the corresponding perturbations.
In the case of the spin relaxation, the coefficient $A$ in Tab. \ref{tab:one} 
will be approximately equal to these strengths divided by a typical energy
difference between the corresponding coupled states, as can be seen from Eq.
\eqref{eq:eigenfunction}. On the other hand, since the excited orbital and
ground states
have the same spin, the coefficient $A$ for the orbital relaxation is of order
1.

This consideration leads to the following approximations which we use when 
estimating the rate analytically. For orbital relaxation 
\begin{equation}
H_1\approx 0.
\label{eq:h1 orbital}
\end{equation}
For spin relaxation, in analytical calculations we neglect the cubic 
Dresselhaus 
term. If the magnetic field is in-plane, Eqs. 
\eqref{eq:hamiltonian2a} and \eqref{eq:hamiltonian2b} do not couple 
the ground or the spin state, which have zero orbital momenta, with any other 
state. If the magnetic field is perpendicular, Eqs. 
\eqref{eq:hamiltonian2a} and \eqref{eq:hamiltonian2b} commute with the 
Zeeman term, again giving no contribution to the spin relaxation. For the spin 
relaxation we therefore approximate
\begin{equation}
H_1\approx H^Z_{D}+H^Z_{BR}.
\label{eq:h1 spin}
\end{equation}

\subsection{In-plane magnetic field}

We have earlier\cite{stano2005:CM} compared our calculation of the spin 
relaxation in an in-plane magnetic field with the experiment.\cite{elzerman2004:N}
We have found that the experiment can be explained with a reasonable set of 
spin-orbit parameters which we use also in the present article.
Using Eqs. \eqref{eq:eigenfunction}, \eqref{eq:rate pz}, and \eqref{eq:h1 spin}
we get for the dominant contribution to the spin relaxation 
due to piezoelectric transversal phonons
in the low magnetic field limit 
\begin{equation}
\Gamma^\mathrm{pz}_t\approx \frac{256\pi \gamma_\mathrm{pz} c_l^2 m^2}{105
\hbar^7 c_t^5} l_{0}^8 |\mu B_\parallel|^5
\mathcal{L}^{-2}_{\mathrm{SO}},
\label{eq:in-plane}
\end{equation}
where
\begin{equation}
\mathcal{L}^{-2}_{\mathrm{SO}}=\frac{
l_{D}^2+l_{BR}^2-2\sin(2\gamma)l_{D}
l_{BR}}{l_{D}^2 l_{BR}^2}
\label{eq:angular}
\end{equation}
describes the effective (anisotropic) spin-orbit length. 
The angular dependence of the spin relaxation rate, expressing the $C_{2v}$ 
symmetry of the heterostructure, allows to find the ratio of the Dresselhaus
and Bychkov-Rashba couplings:
\begin{equation}
\mathrm{min}\{l_{D}/l_{BR},l_{BR}/l_{D}\}=2/(\sqrt{r_a}+1)-1,
\label{eq:ratio}
\end{equation}
where $r_a$ is the ratio of the rates at $\gamma=45^\circ$ and
$\gamma=135^\circ$. 
A possible measured angular dependence with the minimum at $\gamma=45^\circ$
would be a convincing indication that the admixture due to spin-orbit is the
mechanism of the relaxation. A more general angular dependence, allowing
for out-of-plane magnetic fields, was derived
in Ref. \onlinecite{golovach2004:PRL}.

The reason for the angular dependence of $\Gamma_t^{\rm pz}$ 
follows from Eq. \eqref{eq:h1 spin}, which 
for an in-plane field is 
\begin{equation}
H_1=-\mu B_\parallel \sigma_z[
x(\frac{\cos\gamma}{l_{BR}}-\frac{\sin\gamma}{l_{D}})+y(\frac{\sin\gamma}{l_{BR}}
-\frac{
\cos\gamma}{l_{D}})].
\label{eq:anisotropy inplane}
\end{equation}
Due to the selection rules for the Fock-Darwin states, $x$ and $y$ do not mix in
coupling of the states. The coefficient $A^2$
is then proportional to the sum of the squared couplings from Eq.
\eqref{eq:anisotropy inplane}, at $x$ and $y$. 
Taking $E_0$ as a typical energy difference $E$ of the coupled states and
using $l_B$ for a natural length unit, we get
$A^2\approx|\mu B_\parallel l_B/ E_0|^2\mathcal{L}_{\mathrm{SO}}^{-2}$. Noting that
$l_B=l_{0}$ for in-plane field and using the low energy limit for $\Gamma^\mathrm{pz}_t$
from Tab. \ref{tab:one}, one recovers Eq. \eqref{eq:in-plane} 
up to a numerical factor. The numerical result has been presented in Ref. 
\onlinecite{stano2005:CM} (Fig. 1) and is not repeated here.

\subsection{Perpendicular magnetic field}
\subsubsection{Orbital relaxation rates}

\begin{figure}
\centerline{\includegraphics[width=1\linewidth]{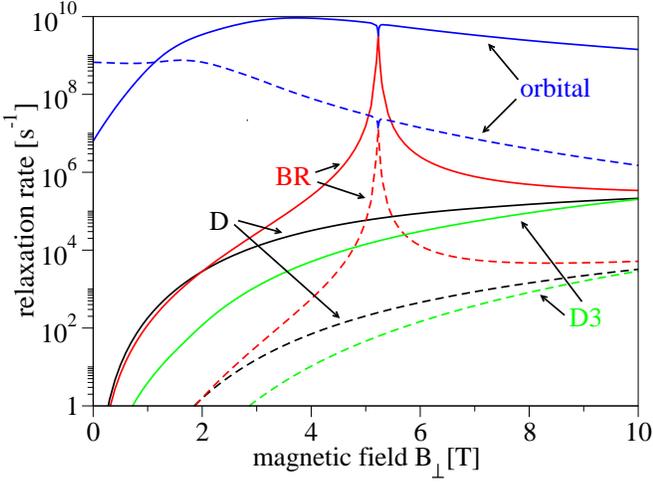}}
\caption{(Color online) Orbital and spin (labels $D,BR,$ and $D3$ denote which 
spin-orbit interaction is present) relaxation rates in a single
quantum dot, for the piezoelectric (solid) and
deformation potential (dashed) phonons. The confining length is 32 nm. Anti-crossing of
the unperturbed spin and orbital state occurs at $B_\perp=5.2$ T.}
\label{fig:sd simple}
\end{figure}

\begin{figure}
\centerline{\includegraphics[width=1\linewidth]{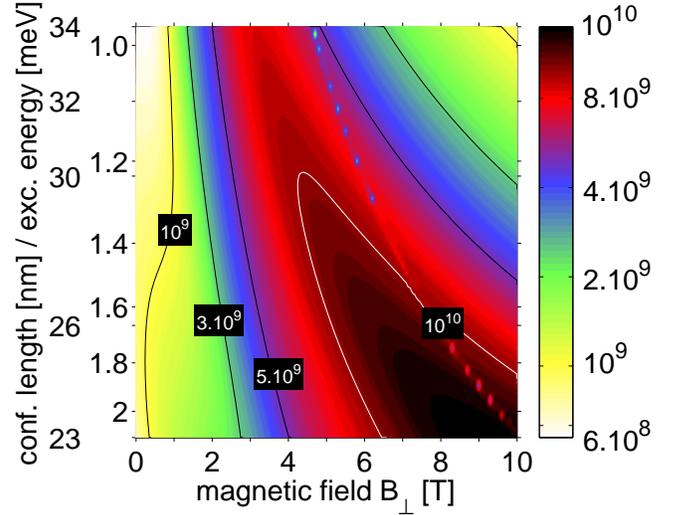}}
\caption{(Color online) Orbital relaxation rate (the sum of the deformation
and piezoelectric contribution) in a single quantum dot `as a function of magnetic field and
the confinement length $l_0$ / the confinement (excitation) energy $E_0$. The rate 
is given on the logarithmic scale in the units of s$^{-1}$. 
The solid lines represent equal relaxation rates
(equirelaxation lines). The granular structure in the figure is an artifact of
the limited data resolution.}
\label{fig:comprehensive orbital single}
\end{figure}

In the case of a perpendicular magnetic field, the numerically
calculated orbital and spin relaxation
rates in a single dot are shown in Fig. \ref{fig:sd simple}.  
The orbital relaxation rate is of the order of $10^9$ s$^{-1}$. 
The spin-orbit contributions to the rate (not shown in the figure) are of the
order of $10^6$ s$^{-1}$ for the linear spin orbit terms and $10^5$ s$^{-1}$ for
the cubic
Dresselhaus term, validating the approximation Eq. \eqref{eq:h1 orbital}. 
\begin{table}
\begin{tabular}{|c|c|c|c|c|}
\hline
\multirow{6}{*}{\begin{sideways}low mag. field\end{sideways}}
&\multirow{3}{*}{\begin{sideways}orbital\end{sideways}}
&$\Gamma^\mathrm{df}$&$(\pi \gamma_\mathrm{df} \hbar^3
c_l/ m)l_{0}^{-4}(1-B_\perp e l_{0}^2 /2\hbar)$&$\lesssim 0.6$ T\\
&&$\Gamma^\mathrm{pz}_l$&$(459\pi\gamma_\mathrm{pz} c_l^5  m^5 /4 
\hbar^5)l_{0}^{4}(1+5 B_\perp e l_{0}^2/2\hbar)$&$\lesssim 0.5$ T\\
&&$\Gamma^\mathrm{pz}_t$&$(61\pi \gamma_\mathrm{pz}
c_l^2 c_t m^3 /4\hbar^3)l_{0}^{2}(1+3 B_\perp e l_{0}^2/2\hbar)$&$\lesssim 0.8$ T\\
\cline{2-5}
&\multirow{3}{*}{\begin{sideways}spin\end{sideways}}
&$\Gamma^\mathrm{df}$&$(128\pi\gamma_\mathrm{df} m^2 /3
\hbar^7  c_l^3) l_{0}^8 |\mu B_\perp|^7 l_{D}^{-2}$&$\lesssim 4$ T \\
&&$\Gamma^\mathrm{pz}_l$&$(128\pi \gamma_\mathrm{pz} m^2 /35
\hbar^7 c_l^3) l_{0}^8 |\mu B_\perp|^5 l_{D}^{-2}$&$\lesssim 4$ T\\
&&$\Gamma^\mathrm{pz}_t$&$\Gamma^\mathrm{pz}_l\times 4c_l^5/3c_t^5$&$\lesssim 4$ T\\
\hline
\multirow{6}{*}{\begin{sideways}high mag. field\end{sideways}}
&\multirow{3}{*}{\begin{sideways}orbital\end{sideways}}
&$\Gamma^\mathrm{df}$&$(2\pi \gamma_\mathrm{df} \hbar^{13}
/3 e^6 m^5 c_l^3)l_{0}^{-20} B_\perp^{-6}$&$\gtrsim 4$ T\\
&&$\Gamma^\mathrm{pz}_l$&$(8\pi\gamma_\mathrm{pz}\hbar^7/35
e^4  m^3  c_l^3) l_{0}^{-12} B_\perp^{-4}$&$\gtrsim 4$ T\\
&&$\Gamma^\mathrm{pz}_t$&$\Gamma^\mathrm{pz}_l\times 4c_l^5/3c_t^5$&$\gtrsim 6$ T\\
\cline{2-5}
&\multirow{3}{*}{\begin{sideways}spin\end{sideways}}
&$\Gamma^\mathrm{df}$&$(32\pi\gamma_\mathrm{df}
|{\mu}|^5/3 \hbar e^2 c_l^3) B_\perp^3 l_{D}^{-2}$&$\gtrsim 8$ T\\
&&$\Gamma^\mathrm{pz}_l$&$(32\pi\gamma_\mathrm{pz} |{\mu}|^3
/35 \hbar e^2  c_l^3)B_\perp l_{D}^{-2}$&$\gtrsim 7$ T\\
&&$\Gamma^\mathrm{pz}_t$&$\Gamma^\mathrm{pz}_l\times 4c_l^5/3 c_t^5$&$\gtrsim 7$ T\\
\hline
\end{tabular}
\caption{Approximate orbital and spin (due to Dresselhaus coupling)
relaxation rates in a single quantum dot at low and high magnetic 
fields in lowest order of the non-degenerate perturbation theory. In the last 
column we state the maximal or minimal magnetic field by requiring that at 
$l_{0}=32$ nm the presented approximation
does not differ from the numerical value by more than a factor of 2.}
\label{tab:two}
\end{table}
The energy difference of the orbital and the ground state is
$E_{}=\hbar^2/2m l_B^2-(\hbar e/2m) B_\perp$. At {\it low}
magnetic fields the {\it high} $\mathcal{E}$ limit applies
and the deformation potential dominates the orbital relaxation rate. 
The results are listed in Tab. 
\ref{tab:two}. The values at zero magnetic field, up to a numerical factor, 
follow from Tab. \ref{tab:one}, if one uses $A=1$ and the low magnetic field 
limits, where $E_{}\approx\hbar^2/m l_0^2$, and $l_B\approx l_0$. 
The dependence of the rates on the energy difference of the states, shown in Tab. \ref{tab:one}, is 
enough to understand the different dependence of the 
deformation and piezoelectric contributions to the orbital relaxation 
rate at low magnetic fields shown in Figs. \ref{fig:sd simple} and \ref{fig:comprehensive orbital single}.
The deformation contribution drops with increasing both the magnetic field and 
confinement lengths, while the piezoelectric contribution increases with increase 
of these two parameters. 

For fields lower that 1 T the dominant deformation 
contribution manifest itself on Fig. \ref{fig:comprehensive orbital single}. 
At magnetic fields higher than 1 T the piezoelectric contribution dominates. Up 
to about 4 T we are still in the regime $\mathcal{E}\gg1$ and the rate 
grows with increasing magnetic field and increasing confinement length.
Since the 
energy difference $E_{}$ drops with increasing magnetic field,
for magnetic fields $\gtrsim 6$ T we get into the limit $\mathcal{E}\ll 1$. 
The corresponding orbital relaxation rates in \ref{tab:two} then follows from 
Tab. \ref{tab:one} using $A=1$
and high magnetic field limits, where $E_{}\approx\hbar^3/e m B l_{0}^4$ 
and $l_B^2\approx2\hbar/e B$.
This leads to a much stronger drop of the deformation contribution to the rate 
with the increase of both magnetic field and the confinement length, than is the drop 
of the piezoelectric contribution. 

Finally, we explain the influence of the anti-crossing on the orbital relaxation 
rate, seen in Fig. \ref{fig:sd simple}. 
The anti-crossing contributes by an overall factor of $|\beta_{ii}|^2$, see Eq. 
\eqref{eq:eigenfunction2}, which multiplies the orbital relaxation rates listed in Tab. 
\ref{tab:two}. Solving the appropriate secular equation, we get\cite{bulaev2005:PRB}
\begin{equation}
|\beta_{ii}|^2=\frac{1}{2}+\frac{E}{2\sqrt{E^2+|C|^2}},
\label{anticrossing solution 1}
\end{equation}
where $E=\hbar^2/2m l_B^2-(\hbar e/2m) B_\perp-2\mu B_\perp$ is the energy 
difference between the crossing states, and $C=-\frac{\hbar^2}{m l_B 
l_{BR}}(1-B_\perp e l_B^2 /2\hbar)$ is the strength of the coupling between these 
states due to the Bychkov-Rashba term. Away
from anti-crossing $\beta_{ii}\approx 1$, while directly at the anti-crossing the rate is reduced by a factor of 2.
The anti-crossing region for the orbital
relaxation is rather narrow ($\sim$ 0.1 T) and manifests itself as a 
narrow line of the suppression of the rate in Fig. \ref{fig:comprehensive orbital single}.

\subsubsection{Spin relaxation rates}

For spin relaxation the relevant energy difference is the Zeeman splitting, 
$E_{}\approx 2|\mu B|$. Therefore the low energy limit, $\mathcal{E}\ll 1$, 
applies up to rather high magnetic fields ($\sim$10 T). 
Piezoelectric transversal phonons dominate the rate.
The linear spin-orbit terms dominate over the
cubic Dresselhaus term, although the difference becomes smaller for higher 
magnetic fields. We use an example of the linear Dresselhaus term for
analytical expressions.
Using Eq. \eqref{eq:h1 spin} and the limits of low and high magnetic
fields we present the analytical spin relaxation rates in Tab. \ref{tab:two} 
(these results were also derived in Refs. \onlinecite{bulaev2005:PRB} and \onlinecite{khaetskii2001:PRB}).
These formulas approximately follow from Tab. \ref{tab:one}
using $A=|\mu B_\perp|l_B/l_{D} \delta E$ and $E_{}=|\mu B_\perp|$, while
noting that $\delta E=E_0$ for low and $\delta E=|\mu B_\perp|$ for high
magnetic fields. 
The trends described by the Dresselhaus contribution can be seen
in Fig. \ref{fig:sd simple}. The spin relaxation rate grows much steeper with increasing magnetic 
field at low $B_\perp$ (fifth power) than at high $B_\perp$ (first 
power). Interestingly, at high magnetic fields the rate does not depend on the 
confining length.

\begin{figure}
\centerline{\includegraphics[width=1\linewidth]{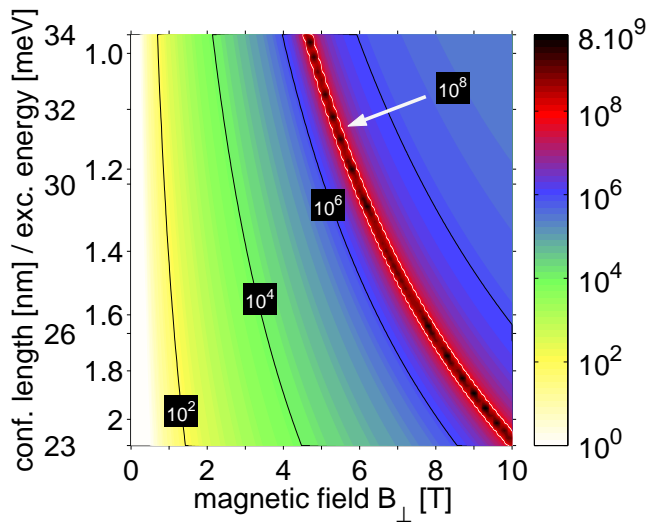}}
\caption{(Color online) 
Spin relaxation rate
in a single quantum dot as a function of magnetic field and
the confinement length $l_0$ / the confinement energy $E_0$. The rate 
is given on the logarithmic scale in the units of s$^{-1}$. 
The solid lines represent equirelaxation lines.}
\label{fig:comprehensive spin single}
\end{figure}

Away from anti-crossing analogous formulas, up to a numerical factor, 
as those listed in Tab. \ref{tab:two}, hold for the contribution to the spin 
relaxation due to the Bychkov-Rashba term after the substitution $l_D \to l_{BR}$.
In this case the contribution to the overlap between the spin and ground states 
due to the term $\beta_{ij}$ in Eq. \eqref{eq:eigenfunction2} is negligible.
However, comparing the analytical formulas from Tab. \ref{tab:two} with the
numerical calculation in Fig. \ref{fig:comprehensive spin single}, 
we find discrepancy, except at low magnetic fields. This is because, as can 
be seen also in Fig. 
\ref{fig:sd simple}, the rate is actually dominated by a spin hot spot (anti-crossing). The anti-crossing 
occurs for single dots only when the Bychkov-Rashba term is present, since the
Dresselhaus terms do not couple the unperturbed orbital states.\cite{bulaev2005:PRB,stano2005:PRB}  
In this case we can neglect all terms but that one containing $\beta_{ij}$ in Eq. \eqref{eq:eigenfunction2} and for the spin relaxation rate due to the anti-crossing one gets $\Gamma({\rm spin, acr})=|\beta_{ij}|^2 \Gamma({\rm orbital})$. The secular equation gives 
\begin{equation}
|\beta_{ij}|^2=\frac{1}{2}-\frac{E}{2\sqrt{E^2+|C|^2}},
\label{anticrossing solution 2}
\end{equation}
where the variables are those defined under Eq. \eqref{anticrossing solution 1}.
Thus, the anti-crossing effectively mixes what we
usually call spin and orbital rates.
The spin relaxation rate has a sharp peak at the anti-crossing. With increasing the 
``distance'' from the anti-crossing the rate drops, mirroring the drop of the 
coefficient $|\beta_{ij}|^2$. Only far enough from the anti-crossing the term 
$\beta_{ij}$ is negligible in Eq. \eqref{eq:eigenfunction2} and the rate is 
described by expressions analogous to those from Tab. \ref{tab:two}. In Fig. 
\ref{fig:sd simple} the Bychkov-Rashba contribution to the spin relaxation rate 
is dominated by the $\beta_{ij}$ term unless the magnetic field is smaller that 
2 T. Similarly in Fig. \ref{fig:comprehensive 
orbital single}, for fields higher than 2 T the total spin relaxation rate is 
dominated by the anti-crossing 
contribution due to Bychkov-Rashba term. Consequently, the influence of the
anti-crossing is substantial in a much larger region (several Tesla) than in the
case of the orbital relaxation. 

In Ref. \onlinecite{bulaev2005:PRB} spin relaxation rates
due to the deformation potential were computed in the lowest order of the
perturbation theory and an analogous figure to our Fig. \ref{fig:sd simple}
was presented. Our results for both orbital and spin relaxation rates are
in a quantitative agreement.

\section{Double dots}

In our double dot potential the ground (excited orbital) state can be
approximated as a symmetric (antisymmetric)
combination of two Fock-Darwin functions, $\Psi_{0,0,\uparrow}$, placed at the
two potential minima.
In Ref. \onlinecite{stano2005:PRB} we have studied the energy spectrum and 
classified the symmetries of the states of a double dot with a potential given by Eq. 
\eqref{eq:doubledot}. What we call here ground, spin and orbital state is denoted 
there as $\Gamma_S^\uparrow,\,\Gamma_S^\downarrow,$ and $\Gamma_A^\uparrow$,
respectively.
The upper index indicates spin and the lower index indicates 
the symmetry of a particular state with respect to spatial inversion. 
The energy difference between the ground and excited orbital state, 2$\delta
E_t$, is strongly influenced by the ratio of the interdot
distance and the effective length,\cite{stano2005:PRB} $\mathcal{D}=d/l_B$,
\begin{equation}
2\delta E_t=\frac{\hbar^2}{m
l_B^2}\frac{2\mathcal{D}(1-\theta^2)\{1+\mathcal{D}\sqrt{\pi}\mathrm{Erfc}
(\mathcal{D})-e^{-\mathcal{D}^2}\} } {\sqrt{\pi} \{e^{\mathcal{D}^2(1+\theta^2)}-e^{-\mathcal{D}^2(1+\theta^2)}\}},
\label{eq:energy dd orbital}
\end{equation}
where a dimensionless parameter $\theta=B_\perp e l_B^2/2\hbar$. 
The tunneling energy $\delta E_t$ gives the frequency of single-electron
coherent oscillations between the left and right dots.  
The approximation of Eq. \eqref{eq:h1 spin} for spin relaxation is
correct also here, since 
Eqs. \eqref{eq:hamiltonian2a} and \eqref{eq:hamiltonian2b} do not couple 
any two of the ground, spin, and orbital states due to 
a definite symmetry of the $L_z$ operator.\cite{stano2005:PRB}  
There is a coupling through higher excited states with appropriate symmetry, 
but, as we learn from numerics, this is negligible.

\subsection{In-plane magnetic field}

\begin{figure}
\centerline{\includegraphics[width=1\linewidth]{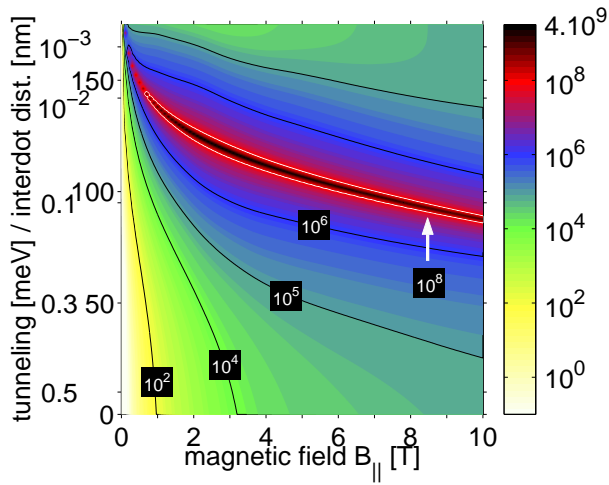}}
\caption{(Color online) Spin relaxation rate in a double quantum dot
as a function of in-plane magnetic 
field for $\gamma=0^\circ$ and the interdot distance $d$ / tunneling energy 
$\delta E_t$, for a confinement length 32 nm.  The relaxation rate is given on
the logarithmic scale in the units of s$^{-1}$. The double dot is oriented along 
[100] ($\delta=0^\circ$).}
\label{fig:comprehensive in-plane double}
\end{figure}

The spin relaxation rate as a function of in-plane magnetic field and the interdot
distance is plotted in Fig. \ref{fig:comprehensive in-plane double}. The rates for 
small  interdot distances are similar to the single dot case, where the
rate grows with increasing magnetic field; for low magnetic fields more steeply 
than for large. The order of magnitude of the rate is given by Eq. 
\eqref{eq:in-plane}, being
about $10^2$ s$^{-1}$ at 1 T and 10$^5$ s$^{-1}$ at 10 T. At large 
interdot distances the rate is strongly influenced by the presence of a
anti-crossing (spin hot spot), which occurs when the Zeeman and twice the tunneling energies are equal.\cite{stano2005:CM} 
If the tunneling energy is changed from zero 
to a value of order of the single dot excitation energy, regardless of the 
magnetic field strength, one always passes through a spin hot spot region, where the 
spin relaxation is very fast. 
Fortunately there exist specific orientations of the double dot system and the 
magnetic field, where this anti-crossing does not occur. We call such a 
configuration ``easy passage.''\cite{stano2005:CM} 

To understand the angular dependence of the rate presented in Ref. 
\onlinecite{stano2005:CM} and find conditions for an easy 
passage we transform the Hamiltonian
\eqref{eq:transformation} with $H_1$ given by Eq. \eqref{eq:h1 spin}
into coordinates in which the new x axis lies along the dot's axis
$\mathbf{d}$. Since there are no orbital effects in in-plane
magnetic fields, in these new coordinates 
the unperturbed solutions of the Hamiltonian $H_0$ have a definite symmetry under
inversions about $\hat{x}$ -- the ground and spin states are symmetric, while the
orbital state is antisymmetric. The transformed $H_1$ of Eq. \eqref{eq:h1 spin}, is
\begin{equation}
 \begin{split}
H_1=&-\mu B_\parallel\sigma_z
\large\{x[l_{BR}^{-1}\cos(\gamma-\delta)-l_{D}^{-1}\sin(\gamma+\delta)]\\
&+y[l_{BR}^{-1}\sin(\gamma-\delta)-l_{D}^{-1}\cos(\gamma+\delta)]\large\}.
\label{eq:anisotropy inplane 2}
\end{split} \end{equation}
In the single dot case the coefficient
$A^2$ in Tab. \ref{tab:one} is
proportional to the sum of the squared couplings in Eq. \eqref{eq:anisotropy inplane 2} at
$x$
and $y$. However, in the double dot case, $x$ and $y$ can couple states
differently. For large interdot distances the most important influence on the
spin relaxation comes from the anti-crossing of the spin and orbital states,
which are coupled by terms with the x-like symmetry. 
Thus, the anti-crossing will not occur if
\begin{equation}
l_{BR}^{-1}\cos(\gamma-\delta)-l_{D}^{-1}\sin(\gamma+\delta)=0.
\label{eq:passage}
\end{equation}
The angles $\gamma$ and $\delta$ that satisfy the above equation define an easy 
passage. For a
double dot oriented along [100] direction ($\delta=0$) the
easy passage occurs for an in-plane magnetic field oriented along
angle $\gamma$ given by 
$\tan \gamma=l_{D}/l_{BR}$.
Similarly to the single dot case, the measured angular dependence
recovers the ratio of the spin-orbit
couplings. Now also revealing which one is larger. More important,
as can be seen from Eq. \eqref{eq:passage}, both linear Bychkov-Rashba and
Dresselhaus (also cubic) spin-orbit terms 
contribute to the anti-crossing; in single dots it is only the Bychkov-Rashba
coupling which gives relevant spin hot spots. The position of the easy 
passage is then given by an interplay of all the spin-orbit terms.
If the double dot is oriented along [110] ($\delta=\pi/4$), the condition for the easy 
passage is $\gamma=135^\circ$, being independent on the spin-orbit couplings. 
The importance of this result has been pointed already in Ref. \onlinecite{stano2005:CM},
where the corresponding numerical results are presented.

\subsection{Perpendicular magnetic field}

\subsubsection{Orbital relaxation rate}

There are two different regimes for the orbital relaxation, 
depending on the energy difference of the ground and orbital states, $E=2\delta E_t$, which is more sensitive to 
the interdot distance than to the confinement length.
If $\mathcal{D}\equiv d/l_B\ll 1$, then $E\approx \hbar^2/ m l_B^2$, 
decreasing with increasing the magnetic field or the interdot distance.
The limit of high
$\mathcal{E}$ applies and the rates are comparable to the single
dot case. On the other hand, if $\mathcal{D}\gg 1$ the energy, and thus also the
rates, drop exponentially with increasing the magnetic field or the interdot distance. Due
to the complex interplay of the magnetic field and interdot distance, no power law
dependence of the rates on magnetic field can be identified. However,
approximations in Tab. \ref{tab:one} give analytical formulas with a fair agreement with numerics, if the energy
difference $E \approx 2\delta E_t$ which is given by Eq. \eqref{eq:energy dd
orbital}.
\begin{figure}
\centerline{\includegraphics[width=1\linewidth]{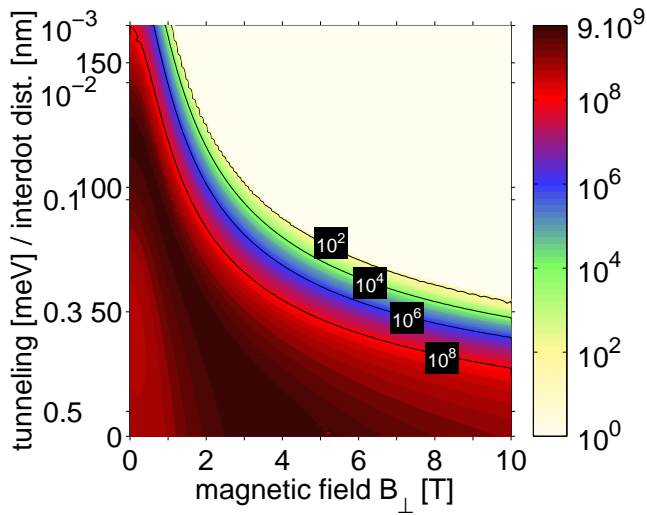}}
\caption{(Color online) Orbital relaxation rate in a double
quantum dot as a function of in-plane 
magnetic field for $\gamma=0^\circ$ and the interdot distance $d$ / tunneling 
energy  $\delta E_t$, for a confinement length 32 nm.  
The relaxation rate is given on the
logarithmic scale in the units of s$^{-1}$. The double dot is oriented along 
[100] ($\delta=0^\circ$).}
\label{fig:comprehensive orbital double}
\end{figure}

The dependence of the orbital relaxation rate on the magnetic field and the
interdot distance,  for a confining length 32 nm, is shown in Fig.
\ref{fig:comprehensive
orbital double}. The lower left corner is the regime of the high $\mathcal{E}$
limit.
The rate here is similar to the single dot case. The opposite corner
is the regime of an exponentially small energy difference and the rate is
practically zero. The transition
between these two regimes comes for a smaller
interdot distance if the magnetic field is higher, since the transition occurs
when $d\sim
l_B$. Again, as in the single dot case, the anti-crossing does not have a large
influence on the orbital rate -- in the figure it can hardly be seen. For
interdot distances much larger than $l_B$ the dots are
effectively isolated.

\subsubsection{Spin relaxation rate}

Spin relaxation in double dots reveals a surprising complexity
as compared to the single dot case. The complexity is due to the
strong anisotropy of spin hot spots. While anisotropy appears
already in single dots, caused by the interference of the
Bychkov-Rashba and Dresselhaus couplings, additional anisotropy
appears in spin hot spots. This anisotropy does not require
the presence of both couplings. Instead, it is caused by the
selection rules for spin-orbit virtual transitions in the 
double-dot spectrum. The corresponding physics is described
by the transformed Hamiltonian $H_1$ of Eq. \ref{eq:anisotropy inplane 2}.
We have presented the corresponding numerical calculation 
in Ref. \onlinecite{stano2005:CM}. Here we discuss the individual
contributions of the Bychkov-Rashba and Dresselhaus terms in 
the spin relaxation rate and, specifically, in the spin
hot spot anisotropy.

\begin{figure}
\centerline{\includegraphics[width=0.8\linewidth]{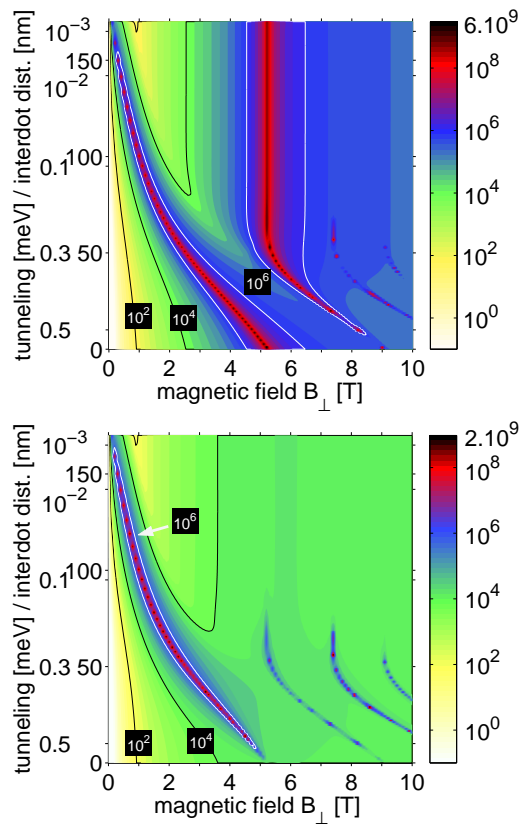}}
\caption{(Color online) Spin relaxation rate as a function of perpendicular 
magnetic  field for $\gamma=0^\circ$ and the interdot distance $d$ / tunneling 
energy $\delta E_t$ (at zero magnetic field only), for a confinement length 32 
nm.  
The relaxation rate is given in logarithmic scale in the units of s$^{-1}$. The 
double dot is oriented along [100] ($\delta=0^\circ$). The upper figure shows 
results when only the Bychkov-Rashba term is present in the Hamiltonian. In the 
lower figure, only the Dresselhaus terms are present.}
\label{fig:comprehensive spin double}
\end{figure}

The contribution to the spin relaxation rate from the Bychkov-Rashba 
(Dresselhaus) term is shown in the upper (lower) part
of Fig. \ref{fig:comprehensive spin double}. The changes of the upper
figure, if the Dresselhaus terms were present, would be very small (compare with Fig. 2 in Ref. \onlinecite{stano2005:CM}).
For low magnetic fields the rate grows with increasing magnetic
field, as
we expect from Tab. \ref{tab:one}. However, similarly to the in-plane magnetic 
field case, the spin hot spots (ridges in Fig. \ref{fig:comprehensive spin double})
dominate the rate for most of the parameters' range.  
The interdot distance strongly influences the spin relaxation 
rate by determining the position of anti-crossings. 
In high magnetic fields, the spin state
can anti-cross higher orbital states depending on the symmetry of these states. 
However, the 
influence of these anti-crossings on the rate is
limited to a narrow region of magnetic fields, since the dots are effectively
isolated at high fields and the crossing states do not comply with the selection 
rules for spin-orbit couplings of single dot states.

It is interesting to compare the contribution to the spin relaxation 
by the Bychkov-Rashba and the Dresselhaus terms. Let us first look
at the single dot regime, which in Fig. \ref{fig:comprehensive spin double}
is visible at $d=0$. The spin hot spot appears only for the Bychkov-Rashba
coupling, in line with our earlier observation \cite{stano2005:PRB}.
The Dresselhaus coupling becomes effective only in the coupled-dot
system in which the symmetry of the lowest orbital states allows
the coupling at the level crossings. The coupling is again absent
at two isolated dots ($d\to \infty$). Another nice feature seen in 
Fig. \ref{fig:comprehensive spin double} is the transformation of the
single-dot spin hot spot at about 5 T to a double-dot spin hot 
spot at lower fields, while the single-dot spin hot spot that starts
at about 9 T shifts towards 5 T in the double dot and remains there 
at all couplings.

Similarly to the in-plane field case, we can 
understand the anisotropy of the spin relaxation in perpendicular
magnetic field by transforming the Hamiltonian of Eq. \eqref{eq:h1 spin} into a coordinate
system with the x-axes being along $\mathbf{d}$:
\begin{equation} \begin{split}
H_1=&\mu
B_\perp\large\{x[\sigma_x(l_{BR}^{-1}-l_{D}^{-1}\sin2\delta)-\sigma_y l_{D}^{-1}\cos
2\delta]\\
&+y[\sigma_y(l_{BR}^{-1}+l_{D}^{-1}\sin2\delta)-\sigma_x l_{D}^{-1}\cos
2\delta]\large\}.
\label{eq:anisotropy perpendicular}
\end{split} \end{equation}
Due to the presence of the orbital effects of the perpendicular 
magnetic field, the unperturbed states have no
specific symmetry under inversions along $x$. As a result only in the limit of low
magnetic fields ($l_B\approx l_{0}$), for us below 1 T, the term in
Eq. \eqref{eq:anisotropy perpendicular}  containing $x$ dominates over the term
containing $y$; in higher fields both terms contribute.
In this limit the condition for a suppression of the anti-crossing is
$l_{D}=l_{BR}$ and $\delta=45^\circ$. This we call a ``weak passage'', 
since the anti-crossing, while strongly suppressed, is still present. 
If the condition for a weak passage is not fulfilled, the 
spin relaxation rate, as a function of $\delta$, still has
a minimum at $\delta=45^\circ$ and a maximum at $\delta=135^\circ$.
However, the ratio between the two extremal values is in general of order 1.

\subsection{Other growing directions}

\begin{figure}
\centerline{\includegraphics[width=1\linewidth]{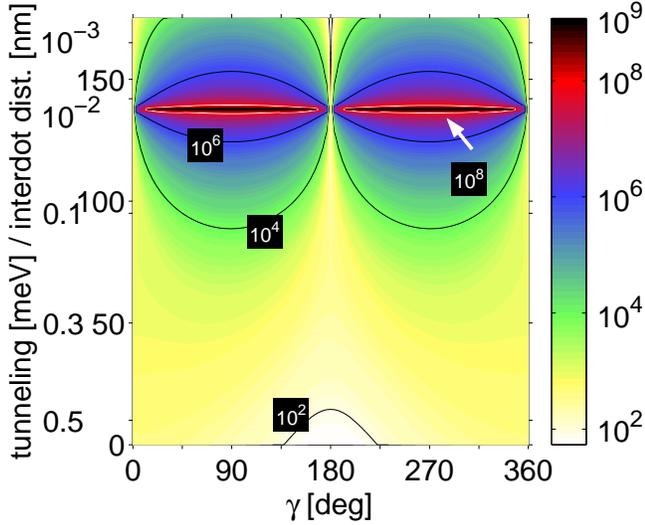}}
\caption{(Color online) Spin relaxation rate as a function of $\gamma$ and the tunneling energy for $B_{||}=1$ T, for [110] growing direction. The dot orientation is given by $\delta=\pi/2$. The relaxation rate is given in logarithmic scale in the units of s$^{-1}$.}
\label{fig:easy passage}
\end{figure}

\begin{figure}
\centerline{\includegraphics[width=1\linewidth]{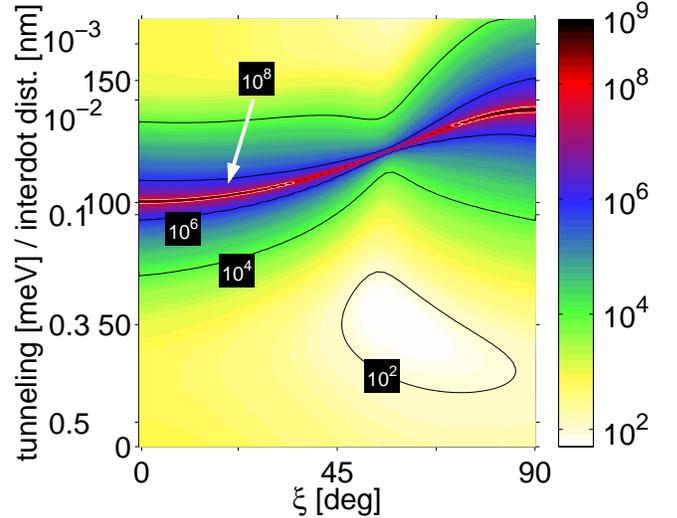}}
\caption{(Color online) Spin relaxation rate as a function of $\xi$ and tunneling energy for $B=1$ T, for [110] growing direction. The dot orientation is given by $\delta=\pi/2$. The relaxation rate is given in logarithmic scale in the units of s$^{-1}$.}
\label{fig:weak passage}
\end{figure}

Thus far we have considered lateral quantum dots defined in 
a (001) plane of a GaAs heterostructure. A different growing direction  
leads to a different form of the Dresselhaus spin-orbit 
interactions\cite{zutic2004:RMP} (the form of the Bychkov-Rashba term remains unchanged) and to different conditions
for the easy passage. Our results are summarized in Tab. 
\ref{tab:passage}. For the [111] growth
direction the Dresselhaus term has the same form as the 
Bychkov-Rashba one. Our results easily translate for this
case by placing formally $l_D \to \infty$. There will be
no spin relaxation anisotropy in single dots, while in 
double dots spin hot spots vanish for $\cos(\gamma-\delta)=0$
at in-plane fields. For a general magnetic field  a
weak passage occurs only at specific spin-orbit parameters,
given by $2\sqrt{3} l_{BR} + l_D =0$ (the couplings can
be negative).

\begin{table}
\begin{tabular}{|c|c|c|}
\hline
growing dir.&in-plane&general\\
\hline
$[001]$&$l_{BR}\cos(\gamma+\delta)=\phantom{eee}$&$l_{D}=l_{BR},\,\delta=\pi/4$\\
&$\phantom{eee}=l_{D}\sin(\gamma-\delta)$&\\
\cline{1-3}
$[111]$&$\cos(\gamma-\delta)=0$&$2\sqrt{3}l_{BR}+l_{D}=0$\\
\cline{1-3}
$[110]$&$\gamma=0,\,\delta=\pi/2$&$l_{BR}\cos\delta=\pm 2l_{D}\cot\xi,$\\
&&$\sin(\delta-\gamma)=\pm 1$\\
\cline{1-3}
$[\cos\alpha\, \sin\alpha\, 0]$ & $\delta=\pi/2,$ &$l_{D}=-l_{BR}\cos2\alpha,$\\ 
&$l_{D} \tan\gamma=-l_{BR}\cos 2\alpha$&$ \delta=\pi/4,\,\xi=0$\\
\hline
\end{tabular}
\caption{Easy passage conditions for several growing directions in an in-plane 
magnetic field and weak passage conditions for a magnetic field with a 
nonzero perpendicular component. The $z$-axis points in the growth direction.
The angle between  $\mathbf{d}$ and the (accordingly rotated) $x$-axis is 
$\delta$, the angle between the in-plane part of the magnetic field and the  
$x$-axis is $\gamma$, while $\xi$ is the angle between the magnetic field and the $z$-axis.}
\label{tab:passage}
\end{table}

A less trivial situation occurs for the [110] grown
quantum well. 
The linear Dresselhaus coupling has the
form
\begin{equation}
H_D = -\frac{\hbar}{4 m l_D} \sigma_z P_x
\end{equation}
Unlike the Bychkov-Rashba coupling, which has eigenspins
always in the plane, the [110] Dresselhaus term has eigenspins
oriented out of the plane.

The calculated spin relaxation rate for
the double dot system oriented along $\delta = \pi/2$ in
an in-plane magnetic field of $B_{||} = 1$ T
is shown
in Fig. \ref{fig:easy passage}. The spin hot spots
exist for all orientations of the field except at
multiples of $\pi$. This is confirmed by analytical 
considerations summarized in Tab. 
\ref{tab:passage}. The easy passage exists if the 
dot is oriented along the (rotated) x axis, while the in-plane magnetic field is 
along $\hat{y}$. Also, the [110] Hamiltonian is not invariant under the 
in-plane inversion of the coordinates which is why the period in $\gamma$ for  the 
relaxation rate is twice as in the case of the [001] growing direction. However, the 
part of the Hamiltonian important for anti-crossing is invariant with respect 
to inversion along $\hat{y}$. Therefore, the results in Fig. \ref{fig:easy 
passage} for $\gamma>\pi$ are equal to those at $2\pi-\gamma$ to a very good 
approximation.

In order to demonstrate the difference between easy and weak passages, 
we plot in Fig. \ref{fig:weak passage} the calculated spin relaxation
rate in double dots defined in a (110) plane. The dots 
are oriented along $\hat{y}$. From Tab. \ref{tab:passage} one gets the conditions for the 
weak passage to be $\gamma=0$, and, for our spin-orbit couplings, $\xi=56^\circ$,
where $\xi$ is the angle between the magnetic field and $\hat{z}$.
This arrangement corresponds to the ``neck'' on the spin hot spot
in Fig. \ref{fig:weak passage}. However, contrary to an easy passage, here the width of 
the anti-crossing region is finite and gets larger with increasing magnetic 
field (not shown).
Since all weak passages we found depend on spin-orbit couplings,
they (better, the corresponding geometries) are much less useful 
for robust inhibiting of spin relaxation than easy passages. 

In the above analysis we have not considered the cubic Dresselhaus term, $H_{D3}$,
in deriving the conditions for easy passages. Being cubic, even after rotating the double dot ($\delta\neq 0$), it always has qualitatively the same symmetry 
properties with respect to inversions about $\hat{x}$ and $\hat{y}$ -- it is a sum of two terms, one with symmetry of $x$ and one $y$. Therefore the
presence of $H_{D3}$ does not destroy the easy passage. It can
only slightly change the conditions for the easy passage to occur. For our
parameters this change, checked numerically, is only on the order of $1^\circ$ for the 
of angles in Tab. \ref{tab:passage}, so the linear terms should provide
a realistic guidance to experimental demonstrations of the predicted
anisotropy.

\section{Conclusions}

We have calculated phonon-induced orbital and spin relaxation rates of single electron
states in single and double quantum dots. The rates were calculated as a function
of in-plane and perpendicular magnetic fields, as well as a function of the 
field and (in the case of double dots) dots' orientation. Realistic, GaAs defined,
electron-phonon piezoelectric and deformation potential Hamiltonians were
considered. Similarly, relevant spin-orbit interactions, namely the Bychkov-Rashba
and linear and cubic Dresselhaus couplings, were used to calculate the spin
relaxation rate. We have supported our numerical findings by analytical
models based on perturbation theory, deriving effective Hamiltonians which 
display, in the lowest order,  all the important effects seen in numerics. 
We have proposed using a classifying dimensionless parameter $\mathcal{E}$ 
which allows to obtain relevant trends and order-of-magnitude estimates
in important limiting cases.

In the case of single dots, we have carefully analyzed the theoretically
predicted anisotropy of the spin relaxation rate in an in-plane
magnetic field. The anisotropy comes from the interplay of the linear
Bychkov-Rashba and Dresselhaus terms (if only one of the terms
dominates, the anisotropy is absent). Experimental verification of the
anisotropy would give a strong evidence of the spin-orbit mechanism 
of spin relaxation. Furthermore, such a measurement would enable to 
estimate the ratio of the two relevant spin-orbit terms. 

For single dots in a perpendicular magnetic field, which causes cyclotron
effects as well as Zeeman splitting, we have numerically investigated the
orbital relaxation rate. In addition, we have provided a simple analytical
scheme to estimate the rates in the important limits of low and high
magnetic fields, and found the corresponding rate as a function of the
confining length. The orbital relaxation rate is found to be of the order 
of $10^9$ s$^{-1}$, with a relatively small dependence on the magnetic field.
At anti-crossings the orbital relaxation rate is reduced by a factor of two.
At low magnetic fields the rate is dominated by the deformation potential
electron-phonon interaction, while at high fields it is dominated by 
piezoelectric phonons. 

On the other hand, the spin relaxation in single dots is always dominated 
by piezoelectric transversal phonons. The contribution of deformation potential
phonons is more than a decade smaller. The rate is on the order of $10^5$ s$^{-1}$ 
over a large region of parameters (magnetic field and excitation energy). 
However, the rate is strongly enhanced in the region of anti-crossing/spin hot spot, 
where it becomes comparable to the orbital relaxation rate. We have also
provided analytical estimates of the rate (away from the spin hot spots)
for various phonon contributions, at the limits of low and high magnetic fields.

\begin{figure}
\centerline{\includegraphics[width=1\linewidth]{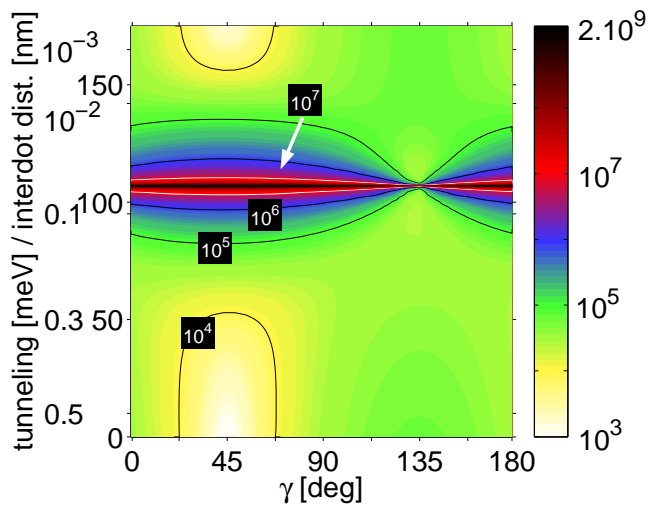}}
\caption{(Color online) Spin relaxation rate in a double dot
as a function of the orientation of 
the in-plane magnetic field and tunneling energy for $B=5$ T, for [001] growing 
direction. The dot orientation is given by $\delta=\pi/4$. A small asymmetric 
term is added into the confinement potential (an electric field of $10^3$ V/m in 
$y$ direction is applied in on one of the dots). By this, the easy passage is 
turned into a weak passage -- compare with Fig. 4 in Ref. 
\onlinecite{stano2005:CM}.}
\label{fig:asymmetry}
\end{figure}

The physics is more complex in coupled dots. We have numerically studied
spin relaxation in double dots in in-plane magnetic fields, in which 
the rate is strongly anisotropic in the direction of both the magnetic 
field and the dots' axis. Similarly to the single dot case, the piezoelectric
phonons dominate spin relaxation here. We have demonstrated that a spin-hot spot exists
at useful magnetic fields (say, 1 T) and interdot couplings (0.1-0.01 meV).
In fact, a spin hot spot is a typical phenomenon in symmetric double dots since
it appears when the tunneling (coupling) energy becomes comparable to the
Zeeman splitting. Fortunately, the spin hot spots are strongly anisotropic,
due to the symmetry of the lowest orbital electronic states, and they
vanish at certain orientations of the field and the dots' axis. We have
systematically investigated these ``easy passages'' using an analytical
model. We have found the criteria for the absence of spin hot spots
for different growth directions of the underlying quantum well. These criteria
should be seriously considered in fabricating double dot systems for 
spin-based quantum information processing which requires low spin relaxation. 

For double dots in a perpendicular magnetic field, the orbital relaxation
rate is most influenced by the energy difference of the corresponding 
coupled states. The energy has a range over eight orders of magnitude due
to the cyclotron effects in the interdot coupling. 
As in the single dot case, both deformation potential
and piezoelectric phonons can dominate the orbital relaxation. 
The spin relaxation in double dots in a perpendicular field has similar qualitative
features as in the single dot case, with an additional anisotropy given by the
orientation of the double dot with respect to the crystallographic axes. 
However, unlike in in-plane fields, only weak easy passages (in which 
spin hot spots form a neck on the parameter map, rather than disappear altogether) exist 
in a perpendicular magnetic field. We have also observed a nice shift of 
spin hot spots to the lower field neighbors as the tunneling between the
dots decreases. While the perpendicular fields provide a nice opportunity
to study fundamental physics of double dot systems, they are less useful
in quantum information processing due to the omnipresence of spin hot
spots and weak passages. 

Our final note concerns the symmetry of the double dot systems investigated
in this paper. Do our conclusions hold if the symmetry is broken? The
answer is yes, if the double-dot system still possesses either x- or
y-like symmetry. Suppose, for example, that a weak electric field is
applied along $\hat{x}$ or $\hat{y}$, or one of the dots is somewhat smaller than 
the other. The spin hot spot anisotropy still leads to easy passages
in spin relaxation in in-plane magnetic fields. On the other hand, 
if the symmetry breaking is $xy$-like (an electric field pointing
along a diagonal, for example), the easy passage is destroyed since
the selection rules for the lowest orbital states will allow coupling
of the states by the term containing $y$ in $H_1$ of 
Eq. \ref{eq:anisotropy inplane 2} (recall that it was the vanishing 
of the term containing $x$ that lead to the appearance of easy passages). This
situation is demonstrated in Fig. \ref{fig:asymmetry}. A double
dot system in an in-plane field of 5 T is oriented along [110]
(the growth direction is [001]). If the double dot is symmetric,
an easy passage exists for $\gamma=135^\circ$ (the corresponding
figure is given in Ref. \onlinecite{stano2005:CM}). However, if one of the
dots is subject to a $y$-like electric field, so that the overall
symmetry of the perturbation is $xy$-like, the easy passage
turns to a weak passage---at all directions of the in-plane
magnetic field there exists an interdot coupling in which 
the spin relaxation rate is greatly enhanced. This is another
important message for spin-based quantum information processing
in quantum dots.

\bibliography{../../references/quantum_dot,../../references/electron-phonon}

\acknowledgments
We thank U. R\"ossler for useful discussions. This work was supported by the US ONR.\\

\end{document}